\documentclass[aps,superscriptaddress,nofootinbib]{revtex4}
\usepackage{amsmath}
\usepackage{graphicx}
\usepackage{epsfig}

\begin{document}



%
%

\title{Quantum coherence to interstellar distances}

\author{Arjun Berera}

\affiliation{School of Physics and Astronomy,
University of Edinburgh, Edinburgh EH9 3FD, United Kingdom}

\begin{abstract}
Quantum coherence could
be sustained up to interstellar distances.
It is shown that the photon mean free path in
certain regions of the electromagnetic spectrum,
such as within the radio or x-ray ranges, could
allow sustaining of the quantum state of a photon
up to galactic distances.
Therefore processes involving quantum entanglement, such as
quantum teleportation, could be realized over
very long distances in the Milky Way or other galaxies.
This is of
fundamental interest and offers a new direction in
the role of quantum mechanics.
Some limited applications of this observation
are discussed.
\end{abstract}


\maketitle


%
%

\section{Introduction}

Quantum teleportation experiments in the last couple of decades
have shown that quantum entanglement can be
maintained over long terrestrial distances.
Such experiments have been done for
photons propagating through
fibre optic cables \cite{mrtzg2003,ujaklwz2004} up to distances of order 10km,
through free space close to sea level, up to around a hundred kilometers
\cite{bbdhp98,fetal2009,yetal2012}, and up to over a thousand
kilometers via satellite to ground teleportation \cite{yetal2017}.
Both through fibre optics and low altitude atmospheric transmission,
loss of signal to the medium has limited the distance to
which teleportation can be successful.  For the satellite based
experiment, where for most of the journey the photons remain
at altitudes above $\sim 10 {\rm km}$, attenuation loss
is substantially less, thus allowing for much longer distance
teleportation.

In quantum teleportation \cite{bennett93}
two photons are entangled \cite{epr35,bell64}, and to
sustain this state, their individual quantum states must be maintained.
Thus long distance entanglement
also means sustaining quantum coherence 
of the individual photons to long distances.
A natural  question arising from the success of the 
long distance terrestrial  quantum teleportation experiments is how
far a distance can quantum entanglement, thus quantum coherence, be sustained.
This paper makes a brief observation that
quantum coherence of photons
can be maintained in some energy ranges to very far interstellar distances
in space.
The primary loss of the photon quantum coherence
in the terrestrial atmospheric free space
teleportation experiments has been
atmospheric turbulence and
other environmental effects like fog, rain, smoke.
These problems are not present in interstellar space, which therefore leaves
the primary loss of quantum coherence to be from
elementary interaction of the photons with other particles
in the medium.
This paper examines the potential interactions of a photon in 
interstellar space
and shows they are weak enough to allow 
a quantum state of a photon to be maintained
to distances from a few parsecs up to the
extent of the Galaxy for certain photon energies.  

\section{Quantum teleportation}

As a simple example of quantum teleportation,
suppose Alice ($A$) and Bob ($B$) are two observers
at different locations, and
Alice possesses a photon
in a quantum state $|\chi \rangle$.
Bob is also in
possession of a photon.
Alice wants to send the complete information about the quantum state
$| \chi \rangle$ over to Bob and input it into
the photon he possesses.  The end result being that
Bob's photon is now in exactly the state $|\chi\rangle$.
If this were achieved, it is
as if Bob has received exactly the photon
Alice had
in her possession.

To implement quantum teleportation, Alice and Bob first need to establish
a shared entanglement with a pair
of photons, say in the Bell state,
\begin{equation}
|\Psi_{AB}^- \rangle = \frac{1}{\sqrt{2}} \left( |+_A -_B \rangle - |-_A +_B \rangle \right) \;,
\label{psiminus}
\end{equation}
where $|+\rangle$ and $|-\rangle$ are the two polarization
states of the photon and
the subscripts $A$ and $B$ correspond to who is in possession of
the respective state, Alice or Bob.
Alice now has the additional photon in the quantum
state, 
$| \chi_{A'} \rangle = c |+_{A'} \rangle + d | -_{A'} \rangle $,
and she wants Bob's photon to be put into this state.
This three particle state then is
\begin{eqnarray}
|\Phi_{A'AB} \rangle & = & c |+_{A'} \rangle |\Psi^-_{AB} \rangle + d |-_{A'} \rangle |\Psi^-_{AB} \rangle
\\
& = & \frac{c}{{\sqrt{2}}} \left[ |+_{A'} \rangle |+_A \rangle |-_B \rangle
- |+_{A'} \rangle |-_A \rangle |+_B \rangle \right]
\nonumber \\
& + &
\frac{d}{{\sqrt{2}}} \left[ |-_{A'} \rangle |+_A \rangle |-_B \rangle
- |-_{A'} \rangle |-_A \rangle |+_B \rangle \right] 
\nonumber \;.
\label{phiaab}
\end{eqnarray}
This state can be reexpressed with the two photons
possessed by Alice written in terms of a Bell state basis, to give,
\begin{eqnarray}
|\Phi_{A'AB} \rangle & = & \frac{1}{2} [
|\Psi^-_{A'A} \rangle (-c | +_B \rangle - d |-_B \rangle)
+|\Psi^+_{A'A} \rangle (-c | +_B \rangle + d |-_B \rangle)
\nonumber \\
& + & |\Phi^-_{A'A} \rangle (c | -_B \rangle + d |+_B \rangle)
+|\Phi^+_{A'A} \rangle (c | -_B \rangle - d |+_B \rangle)
] \;, 
\label{bellform1}
\end{eqnarray}
where the other three Bell states are
$|\Psi^+ \rangle = \frac{1}{\sqrt{2}} \left(|+ - \rangle + |- + \rangle \right)$ 
and
$| \Phi^{\pm} \rangle = \frac{1}{\sqrt{2}} \left( |++ \rangle \pm |-- \rangle \right)$.
Notice that in this change of basis, the states at $B$ now are
all related to $|\chi \rangle$ by a unitary transformation, with the first
of the above terms being in fact exactly the state $|\chi \rangle$ up to
an overall phase of $-1$.  
If Alice now makes an observation of one of
the above four Bell states
at her side, this will collapse the above wavefunction.  Whichever of
the four Bell states she observes, she can communicate that
information to Bob via a classical channel which requires just two 
classical bits of information.
Upon receiving that information, which thus
can arrive no faster than the speed of light, he will know
what state his particle is in.
Thus at this point the state $|\chi \rangle$ up
to an unitary transformation has been teleported
from Alice to Bob.  This process has also
destroyed that state at Alice's side,
consistent with the no-cloning theorem
\cite{wz82}.
As quantum teleportation is a linear operation on
quantum states, this example of
single qubit teleportation can be extended to
teleportation of
multi-particle and multiple degrees of freedom
\cite{ghoshetal2002,vwz2002,lmx2007,dlbpa2011,sp2012,hcth2013,khflrz2014,wcscwlllp2015,drl2016,mehkfz2016}.

To realize quantum teleportation
several technical problems must be overcome.
First, Bell states need to be created, for which
one method is
spontaneous parametric down conversion
\cite{zzhe93,shapiro2002,cbthssag2014}.
Bell states also need to be measured,
for which some techniques have developed
\cite{pewfb2015,bpmewz97}.
For the created Bell state in this process,
both particles of this state need to be
given to the two respective observers $A$ and $B$.  If there are
any environmental factors that interfere with either of
the two particles during
their journey to the two observers,
the quantum coherence between the two
particles will of course be degraded \cite{shapiro2002,betal2014}.
This is the problem to be considered here, from the
demands placed due to the ultra-long distances
required for interstellar propagation of the photons.

\section{Sources of decoherence}

The success of teleportation relies on 
the entangled photons maintaining their individual 
quantum coherence
over the distance between the two participating observers Alice and Bob.
If the entangled photons are transmitted in free space, various
effects from the medium could potentially damage the entangled state
or harm the quantum state engendered on the photons.
For free-space transmission within Earth's atmosphere, turbulence and
other environmental effects like fog, rain, smoke are known to
effect the photons.  This results in absorption of 
the photons, decoherence or phase distortion, all of
which destroy the delicate quantum coherence in the entangled
state \cite{fetal2009,jetal2010,yetal2012,mdt2015}.  Notwithstanding
these potential problems, as already noted  free-space teleportation
has been achieved at sea-level up
to distances of $\sim 100$ km  \cite{fetal2009,yetal2012}
and in ground-to-satellite based tests to
distances of $\sim 1000 {\rm km}$ \cite{yetal2017}.
In the latter, predominant photon decoherence and turbulence occurs
in the troposphere region, within $\sim 10$ km above sea-level.

In interstellar space there are no atmospheric conditions such as fog
and other such effects, 
which are the causes
of entanglement loss in the atmosphere.  That
implies the main source
of entanglement loss and decoherence
in interstellar space would be from elementary interactions of
the entangled photons with particles along their path.
There are regions in the interstellar medium of magnetohydrodynamic
turbulence, but ultimately it is still the elementary interactions of
a photon with the particles in this region that
will lead to decoherence effects on it.

The mean free path of a particle is dependent on
the interaction cross section, $\sigma$, with another particle and
the number density, $n$, of the other particles,
$l_{mfp} = 1/(n \sigma)$. 
Interstellar space 
has a background distribution of hydrogen, electrons,
protons, and there are
photons from the cosmic microwave background (CMB). In addition there
are also some trace amounts of other elements.
The average number density in interstellar space for protons
is about one per ${\rm cm}^3$, giving a mass density
of  $\rho_{ism}= 10^{-21} {\rm kg}/{\rm m}^3$.
For comparison, the mass density of the Earth's atmosphere at 
sea-level is $\rho_a= 1 {\rm kg}/{\rm m}^3$.
There are considerable variations in the proton number density
in the Galaxy, ranging from well below one proton per ${\rm cm}^3$ in
coronal gas regions, to HI regions with around one
per ${\rm cm}^3$, to as high as $10^4$ per ${\rm cm}^3$ in
the HII gas regions, and a few orders of magnitude even higher
in the dense $H_2$ regions \cite{draine2011}.  A large fraction of
this density is ionized with free electrons and protons
and the rest in neutral hydrogen atoms or $H_2$ molecules.

For a first estimate, utilizing the results
of atmospheric entanglement
experiments,
if one assumed the mean free path difference between
the atmosphere and interstellar space simply scales with the density
difference, so ignoring any differences in the specific
particle content in the two systems, then
the mean free path
in interstellar space, using the average
density of one proton per ${\rm cm}^3$,
would scale as $\rho_a/\rho_{ism} \sim 10^{21}$.
Using $\sim 100 {\rm km}$ as the distance that entanglement is
empirically known to sustain in the atmosphere at sea-level,
this would mean in interstellar space it could
sustain entanglement to a distance $\sim 10^{23} {\rm km}$,
which is of order the size of the observable universe.
This is an overestimation because most
of the mass density in Earth's atmosphere is composed
of neutral particles that will not interact so readily with photons,
which interact with charged particles.
In interstellar space, where there would be isolated charged
electrons, protons and ions,
photons would more readily interact.

The interaction of
photons with free electrons or protons at energy below the electron mass is
via Thomson scattering.  This energy range implies photons
in the x-ray region of the spectrum and below.  The Thomson
cross section is,
\begin{equation}
\sigma_{th} = \frac{8 \pi}{3} \left(\frac{\alpha \hbar c}{mc^2}\right)^2 \;,
\end{equation}
where $\alpha \approx 1/137$ is the dimensionless fine structure constant, 
$\hbar$ the Planck constant,
$c$ the speed of light, and $m$ the mass of the
charged particle with which the photon is scattering.
The fine structure constant is a measure of the coupling
in Quantum Electrodynamics (QED) of photons with charged
particles, and as it is much less than one it is
indicative of the weakness of the interaction.
For the electrons in the medium,
this gives a cross section of $6.65 \times 10^{-25} {\rm cm}^2$
and for the proton in the medium, their contribution
is six orders of magnitude smaller.
The mean free path of a photon can now be calculated for 
the different dense regions of the Galaxy.
Using the average value for the number density of free 
electrons or protons in interstellar
space, taking  $n_e \approx 1/{\rm cm}^3$, this leads to
\begin{equation}
l_{th} = \frac{1}{\sigma_{th} n_e} \approx 10^{22} {\rm m} \approx 10^6 {\rm parsec}\;,
\end{equation}
which is longer than the size of the Milky Way Galaxy.
On the other hand, if one
looks at the dense parts of the HII gas region, taking
$n_e \approx 10^4/{\rm cm}^3$, this reduces the mean free path
to $10^2 {\rm parsec}$, which is still traversing a substantial
distance in the Galaxy.

The other background is of the CMB photons.
Their energy density follows from the blackbody expression
as $U_{CMB}= 8 \pi^5/(15 h^3 c^3) (k_B T)^4$.
The presentday CMB photons are at a temperature $2.7 {\rm K}$,
thus the number density of CMB photons
is approximately 
$n_{CMB} \approx U/(k_B 2.7K) \approx 700/ {\rm cm}^3$.
The photon-photon cross section for photons below the electron
mass is \cite{ek35},
\begin{equation}
\sigma{\gamma \gamma} = \frac{973 \alpha^4 (\hbar \omega)^6}{10125 \pi (mc)^8} (\hbar c)^2 \;,
\end{equation}
where $\omega$ is the center-of-momentum frame energy of the
photons and $m$ the electron mass.  In this expression 
$\alpha$ appears to the fourth power, indicating this process
is even weaker than Thomson scattering.
For the CMB temperature 
$2.7K = 2 \times 10^{-4} {\rm eV}$ and for an x-ray photon of
energy $100 {\rm keV}$, that means $\omega \approx 4.5 {\rm eV}$
so $\sigma \approx 10^{-65} {\rm m^2}$.
This implies the mean free path for the x-ray photon due
to interaction with the CMB blackbody photons is,
\begin{equation}
l_{CMB} = 1/(\sigma_{\gamma \gamma} n_{CMB}) \approx 10^{53} {\rm km} \;,
\end{equation}
which is much longer than the observable Universe.
Our interest here is restricted just within the Galaxy, which means
the interaction with CMB photons is negligible.
Electrons and protons, free and bound as hydrogen, as well as photons,
are the main background
constituents prevalent all over the interstellar medium.
The above estimates show that owing to the weakness of
QED, for photons propagating through the interstellar medium
in the energy range of x-rays or lower,
their interaction with this background is negligible.

Gas and dust are also distributed through
the interstellar medium of the Galaxy.
In addition to the dominant
distribution of hydrogen,
there are also trace amounts of other elements
in the interstellar medium, such as helium, carbon, nitrogen,
oxygen, neon, etc...
Photons will also interact with the trace abundances of
these elements through photoabsorption and photoionization.
The expressions
for these interactions are complicated, but
various sources give
the interaction cross sections, opacity and 
the mean free paths in the interstellar medium
\cite{cpbl74,biswas2000,vb2011}.  They find that
the interstellar medium
is transparent to photons in the radio wave region,
energies below $\sim 10^{-3} {\rm eV}$ with some caveats
\cite{gould69}. This continues into the microwave
region. However from the infra-red into the visible and ultraviolet
regions,
the interstellar medium starts to become more opaque due to
the interaction of photons at these energies with atoms
in the interstellar medium. Then above 
tens of ${\rm eV}$ the interstellar medium once again starts to become
increasingly transparent.  In the lower x-ray region at
$100 {\rm eV}$ the photon has a mean free
path around $10 {\rm parsec}$ and for higher energy x-ray photons
at $10^4 {\rm eV}$
the mean free path is above $10^5 {\rm parsec}$, which is
of order the size of the Galaxy.  Thus there
is a wide range of photon energies both in the radio/microwave and then in
the x-ray regions that lead to long mean free paths.

For classical observation this entire range of spectrum can
be detected. However for quantum observations, minimizing
interactions with the interstellar medium will minimize
decohering effects on the delicate quantum coherence the
signal may contain.  For that purpose, the radio and
microwave range and then the x-ray range have advantages.
Magnetic fields are also present in the Galaxy with typical
strength around a $\mu {\rm G}$.  In ionized regions of
the interstellar medium, these magnetic fields affect
the propagation of electromagnetic waves leading to
Faraday rotation and when magnetohydrodynamic turbulence
is present also scintillation. These processes affect
long wavelength electromagnetic signals, so have consequences for
radio waves but are negligible for x-rays \cite{hsw64,rickett77,draine2011}.
In summary, this simple analysis shows that for certain ranges in
the electromagnetic spectrum,
the quantum coherence of an
entangled photon signal could be sustained over vast
interstellar distances.  

\section{Discussion}

This paper has placed focus on the recent successes with
long distance atmospheric quantum teleportation experiments.
These experiments are highly suggestive that much longer
distance teleportation could be possible, and this paper
has explored that possibility.  Using
only known empirical information,
we have been able to deduce that
quantum teleportation and
more generally quantum coherence can be sustained
in space out to vast interstellar distances within the Galaxy.  The main
sources of decoherence in the Earth based experiments,
atmospheric turbulence and
other environmental effects like fog, rain, smoke,
are not present in space.
This leaves only the elementary particle interactions
between the transmitted photons and particles present
in the interstellar medium.  For the most prevalent particles
distributed over the interstellar medium, free electrons,
protons and CMB photons, their interactions with a
propagating photon were computed and extremely long
mean free paths were found.  Other particles in
the interstellar medium have only trace abundances
but can have much stronger interactions
with photons.   Such interactions have been extensively
studied in the literature, and those results can
be transferred over and be applied to the problem
studied in this paper.  Clearly the same reasoning
can apply to examine quantum teleportation and quantum coherence
at intergalactic or cosmological distances as well as 
at energies higher than the
x-ray range.

This paper utilized quantum teleportation as the main example, but
there are many other protocols requiring quantum coherence
to be sustained over spatial distance, such as
quantum key distribution \cite{bb84},
superdense coding \cite{bw92},
and also variants of quantum teleportation such as
remote state preparation \cite{bennettetal2001}.  Alternatively photons
in quantum states could just be individually propagating.
The considerations in this paper would apply to all such cases.

Aside from the energy of the photon, other
factors also dictate the extent of decoherencing
effects on its quantum state.
There can be differences in how quantum coherence
of the individual particles versus the entanglement
between the particles respond to 
decohering effects \cite{wbps2006,ablizetal2006,wwzh2018}.
Specific entangled states have also been shown to respond
differently to decohering effects with some more robust
to withstand these effects \cite{jungetal2008}.
Electromagnetic radiation from astrophysical sources
will be macroscopic, so contain large numbers of photons.
Even though we have shown that photons within certain
energy bands have large mean
free paths through the interstellar medium, just
due to the macroscopically large number of
photons present in any radiation field emitted
from an astrophysical source, some will inevitibly
interact with the medium.  This will lead to incoherent Thomas
scattering events but for some wavelengths could also lead
to identifiable collective behavior such as Faraday rotation
or scintillation.  In the terrestrial quantum teleportation experiments
discussed at the start of the paper,
decoherering effects are a common problem that degrade signal fidelity
and so must be properly accounted for at the receiving end,
when measuring
for the quantum signal.  Similarly
any conceivable  quantum astronomy experiment would have to 
account for decohering effects.  
This decoherence problem is only
further complicated by the fact
the Galaxy is not
homogeneous and isotropic.  Depending on the direction a
signal is being sent, it will experience varying environments
of dust and other features. 
All such factors will be relevant in determining the extent to
which quantum coherence can be maintain on galactic distance
scale.
The main observation made in
this paper is that for certain ranges of photon energies,
the mean free path of such photons is so large in the interstellar medium,
that a large portion of such photons would nevertheless
not decohere.  They would remain
in their initial quantum state at the receiving end 
if they were initially placed
in one at point of emission.

The considerations in this paper are of fundamental interest 
in relation to the role of quantum
mechanics on astrophysical scales.   Immediate application
of these results is limited but there are a few possibilities.
There are some examples of quantum behavior exhibited
by astrophysical bodies. The considerations of this paper
suggest any associated quantum correlations emitted from
these bodies in electromagnetic signals might remain
intact over the long transmission distances in space.
Therefore in addition to any classical signatures,
if such signals retained quantum properties, those might be measurable
with apparatus based on Earth or space based near Earth.

One example of a source that could be producing quantum
coherent signals is the nonthermal radio filament found
near the center of the Milky Way Galaxy,
with one interpretation being it is a light superconducting
cosmic string \cite{mzg2017}. Similar observation of filaments
near the center of the Galaxy have been made before\cite{my85}.
If they were superconducting cosmic strings
\cite{Witten:1984eb,Chudnovsky:1986hc},
they would lead to electromagnetic effects which would have 
macroscopic observables but also underlying quantum signatures

Recently the thermal light from the Sun was used 
to test quantum interference with a photon
sourced in the laboratory 
\cite{dengetal2019,duanetal2020},
with the same test also tried with the nearby extrasolar star Sirius.
This showed that two photons that are sourced at
astronomical distances apart exhibit quantum interference, thus
testing the underlying quantum nature of
photons and their indistinguishability.  The suggestion in
our paper here goes further, that in some cases
the actual quantum states of
photons can be preserved over the long transmission distances of
interstellar space.
A more distant possibility is tests could be done
to probe into the quantum features of Hawking radiation for
primodial blackholes. Such blackholes also have been suggested
can create lasing effects \cite{Rosa:2017ury}, 
for which, beyond its classical electromagnetic
signal, its quantum features could be probed.
All these possibilities suggest a new type of astronomy
looking at quantum features in astrophysical systems.

In a different direction, the results in this paper imply
the quantum communication processes that are
showing success in Earth based tests would also
work and to much greater distance in space.  
For near space applications, such as within the Solar System,
this can already be inferred, but what is more
unexpected is such communication methods could be applicable
even at interstellar distances.  The one possible application
of immediate interest this suggests is attempts at
searching for intelligent extraterrestrial communication signals
could come from a quantum communication mode rather
than the classical communication modes that have
been the only focus up to now.

\acknowledgements
I thank Alan Heavens and Majid Safari for helpful discussions.
This research was supported by the Science Technology Funding
Council (STFC).

\end{document}